\documentclass[aps,prl,reprint,showpacs]{revtex4-1}
\usepackage{wrapfig}
\usepackage{float}
\usepackage[dvips]{graphicx}
\usepackage{subfigure}
\usepackage{amsmath}

\newcommand{\argmax}{\operatornamewithlimits{argmax}}

\begin{document}
\title{Identifying dynamical systems with bifurcations from noisy partial observation}

\author{Yohei Kondo}
\email{kondo@complex.c.u-tokyo.ac.jp}
\author{Kunihiko Kaneko}
\author{Shuji Ishihara}
\affiliation{Graduate School of Arts and Sciences, University of Tokyo, 3-8-1 Komaba,, Meguro-ku, Tokyo 153-8902, Japan}

\date{\today}

\begin{abstract}
  Dynamical systems are used to model a variety of phenomena
 in which the bifurcation structure is a fundamental characteristic.
 Here we propose a statistical machine-learning approach to derive low-dimensional models
 that automatically integrate information in noisy time-series data from partial observations.
 The method is tested using artificial data generated from two cell-cycle control system models
 that exhibit different bifurcations, and the learned systems are shown to robustly inherit
 the bifurcation structure.
\end{abstract}

\maketitle

Various phenomena ranging from climate change to chemical reactions have been modeled extensively
by dynamical systems \cite{Saltzman2001,Strogatz2001},
and the relevance of dynamical systems to modeling biological phenomena is being increasingly recognized
\cite{Kaneko,Alon}.
Recent advances in experimental techniques such as live-cell imaging that clarifies molecular activities
at high spatiotemporal resolutions \cite{Stephens2003,Shav-Tal2004,Fernandez-Suarez2008}
have accompanied this recognition.
However, noise, partial observation, and a low controllability are still challenges for measuring
 biological systems in that both the system dynamics and measurement processes are highly stochastic,
only a few components in a system are observable, and only a small number of experimental conditions
can be examined.
These difficulties have led to models being constructed from experimental observations
in a way that is often ad-hoc and semi-quantitative at best because instructive criteria
 and practical methods have not yet been established for deriving the model equations
by systematically integrating the information in the experimental data.

To model complex systems such as cellular processes,
a full description of all the systems details is often impractical
 and not informative. Instead, a reduced description that preserves the essential features of the system
 is more useful for comprehension., i.e., models described by low-dimensional dynamical systems
 are sufficient for explaining experimental observations.
In particular, the bifurcation structure is a fundamental feature of dynamical systems
 since it characterizes the qualitative changes of the dynamics.
Thus, identification of low-dimensional model systems that inherit the original bifurcation structure
 is a crucial step in understanding the dynamics.

\begin{figure}[b]
  \centering
  \includegraphics[width=8cm]{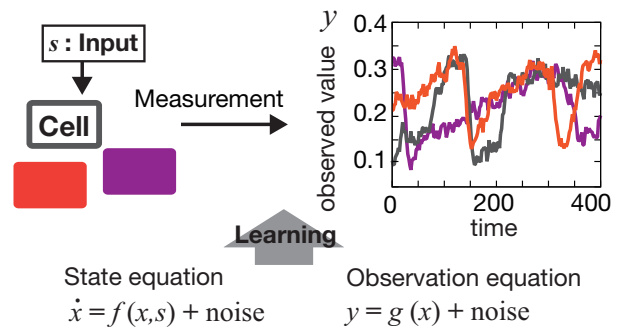}
  \caption{
    Schematic representation of the proposed method.
    Time-series data of the molecular activities in individual cells are obtained from measurements
    under a given input level $s$, which is regarded as a bifurcation parameter of the system.
    System and observation equations are then trained to reproduce the time-series data.
  }
  \label{fig:scheme}
\end{figure}

Here we propose a statistical machine-learning approach to automatically
derive the low-dimensional model equations from single-cell time-series data obtained
at a few conditions (i.e., bifurcation parameter values; Fig. \ref{fig:scheme}).
Techniques for learning nonlinear dynamical systems from time-series data
 have been employed for chaos \cite{Judd1996,Wang2011a},
 spatiotemporal patterns \cite{Voss1999,Muller2002,Sitz2003}, and multi-stable systems \cite{Ohkubo2011}.
Only a few studies have applied the technique to biological data \cite{Nagasaki2006, Yoshida2008}.
In a similar manner to some of those studies \cite{Nagasaki2006,Yoshida2008},
we employ a statistical technique to deal with the noisy and partial time-series data.
However, rather than aiming to fit the model parameters to the observation,
 we obtain the low-dimensional model equations that inherit
 bifurcation structure of the full system to capture the basic nature of the observed system.
The performance of the method is demonstrated by using artificial data.

We introduce a nonlinear state space model composed of state and observation equations
 that describe the system dynamics and observation process, respectively.
 We consider a system that is modeled
 by a $D$-dimensional stochastic differential equations,
 and $d$ components in the model can be simultaneously observed.
The state equations are discretized in time by the Euler-Maruyama scheme \cite{Kloeden}.
 We write the time evolution of the $i$th variable at a time point $t$, $x_i^t (i=1,\dots,D)$, as
\begin{eqnarray}
  x_{i}^{t+1} &=& x_{i}^{t} + \Delta t f_i(\{x_j^t\},s) + \sigma_i\xi_i^t\sqrt{\Delta t},
\end{eqnarray}
where $\Delta t$ is an integration time, $\sigma_i$ is the intensity of the system noise,
and $s$ is a bifurcation parameter.
System noise $\xi_i^t$ is sampled from a standard normal distribution.
To achieve efficient learning, the function $f_i$ is considered to be expressed by a
summation of linearly independent functions as $f_i(\{x_j^t\},s) = \sum_n^{N_i}k_i^nf_i^n(\{x_j\},s)$,
where $N_i$ is the number of parameters $\{k_i^n\}$ and functions $\{f_i^n\}$.
Since our aim is to reproduce the bifurcation structures of systems subjected to unknown equations,
we adopt a polynomial base for the $\{f_i^n\}$,
rather than biochemically realistic functions like the Michaelis-Menten equation.

The observation value of the $i$ th component at a time point $r$, $y_i^r (i=1,\dots,d)$, is written as
\begin{eqnarray}
  y_i^r &=& g_i(x_i^{r}) + \eta_i \phi_i^r
\end{eqnarray}
where $\eta_i$ is an observation noise intensity,
and $\phi_i^r$ is sampled from a standard normal distribution.
In general, a set of observed time points is a part of the entire set of time points in the numerical integration.
Hereafter, $\theta$ indicates the parameters to be estimated:
$\left( \{k_i^n\},\{\sigma_i\},\{\eta_i\}\right)$.

The learning of dynamical systems is formulated as a maximum likelihood (ML) estimation,
which is summarized below (further details are given in the section 1 in the Supplemental Material).
The likelihood is given by the conditional probability of the observed time series $Y$
as a function of the model parameters $\theta$.
However, a straightforward maximization of the likelihood $p(Y|\theta)$ is difficult
because it requires
the untractable summation of $p(Y|X,\theta)p(X|\theta)$ with respect to the time series of the state variables $X$.
Thus, we employ the EM algorithm to maximize the log likelihood of a model by a two-step iterative method
that alternately estimates the states and parameters \cite{Dempster1977}.
In the first step, the E step, the posterior distribution of the time series of a state ($p(X|Y,\theta)$)
is estimated based on the tentative parameter set $\theta_{\rm old}$.
In the second step, the M step, the expectation value of $\log p(X,Y|\theta)$ is calculated as
\begin{eqnarray}
  Q(\theta,\theta_{\rm old}) = \langle \log p(X,Y|\theta)\rangle_{p(X|Y,\theta_{\rm old})}\label{eq:estep},
\end{eqnarray}
and the parameter estimation is updated as
\begin{eqnarray}
  \theta_{\rm new} = \argmax_{\theta} Q(\theta,\theta_{\rm old}).\label{eq:mstep}
\end{eqnarray}
In this step, the optimization problem is reduced to linear simultaneous equations
 and thus can be solved easily.
However, the problem in the E step is still analytically unsolvable
 because the probability distribution of the time series is necessary.
This calculation requires a state estimation at all time points
including the points at which measurements are not conducted.
We therefore obtain a numerical approximation of $p(X|Y,\theta)$ using a particle filter algorithm
 that performs state estimations of nonlinear models using a Monte-Carlo method \cite{Gordon1993, Kitagawa1996}.
The particle filter (a numerical extension of Kalman filter)
approximates a general non-Gaussian state distribution as a set of particles
representing samples from the distribution
and evaluates the log likelihood of the models.
Since the use of the particle filter introduces stochasticity into the learning algorithm,
a slight modification of the M step is required to ensure convergence of the learning \cite{Delyon1999}.
The optimization function in eq.(\ref{eq:mstep})
is replaced by $Q'_{I}(\theta) = (1 - \alpha_I)Q'_{I-1}(\theta) + \alpha_I Q(\theta,\theta_{\rm old})$,
where $I$ is the iteration index, and $\{\alpha_I\}$ is a sequence
of non-increasing positive numbers converging to zero.

To validate the method, we apply it to artificial data
 generated from models of a eukaryotic cell cycle control system
 since this system provides an illustrative example of cellular dynamics
 composed of many molecular components \cite{Novak1993,Borisuk1998,Pomerening2005,Tsai2008,Pfeuty2008}.
The cell cycle is a fundamental biological process characterized by repeated events
 underlying cell division and growth in which key proteins,
 Cyclin and Cyclin-dependent kinases, change their concentration periodically and activate
 various cellular functions such as DNA synthesis.

Two molecular circuit models of the cell-cycle control system in {\it Xenopus} embryos are adopted
 as the data generators:
that proposed by Tyson and co-workers (the Tyson model) \cite{Novak1993,Borisuk1998},
 and that proposed by Ferrell and co-workers (the Ferrell model) \cite{Pomerening2005,Tsai2008}.
Although both models show an oscillation onset as the synthesis rate of Cyclin increases,
they differ in the type of bifurcation at the onset;
the Tyson model exhibits a saddle-node bifurcation on an invariant circle (SNIC),
 while the Ferrell model exhibits a supercritical Hopf bifurcation.
We investigate whether the proposed learning procedure
 reproduces the correct bifurcation types of each model.

Both data generators are composed of 9 variables
 including Cdc2, Cyclin, and other regulatory proteins.
The time-series data is generated by a numerical calculation
 of these models as nonlinear Langevin equations at a few parameter values
(see the section 2 in the Supplemental Material for the model equations, and obtained time-series data).
We simulate noisy observation by adding Gaussian noise to each observation value.
Artificial data are prepared for three Cyclin synthesis rates across the bifurcation point,
and for each bifurcation parameter value, three independent time series are prepared in which
the oscillation exhibits a large fluctuation in amplitude and period among the samples.

Considering a polynomial of degree $M$, we write the system equations to be learned as
\begin{eqnarray}
  f_i(\{x_j^t\},s) = k_i^1 + k_i^2x_1^t + k_i^3x_2^t + \dots + k_{i}^{N_{i}}(x_D^t)^M
\end{eqnarray}
The observation equations are expressed simply as $y_i^r = x_i^{r} + \eta_i \phi_i^r$.
We consider the active Cdc2 and Cyclin concentrations to be observable variables
since their levels have been observed in previous experiments \cite{Pomerening2005}.
Accordingly, $y_1$ $(x_1)$ and $y_2$ $(x_2)$ represent the observed (true) concentrations
 of active Cdc2 and Cyclin, respectively.
The other variables $x_i$ $(i>2)$ represent the true concentrations of unobservable components.
In the system, the Cyclin synthesis rate, $s$, is a bifurcation parameter.
We take the constant term in the equation for Cyclin to be the synthesis rate, i.e., $k_2^1 = s$.
Note that the observed orbit in the active Cdc2-Cyclin plane exhibits no intersection,
(see Fig. S1 in the Supplemental Material)
 suggesting that the two variables are sufficient to abstract the original high-dimensional dynamics.

\begin{figure}[tbp]
  \centering
    \subfigure[]{\includegraphics[width=8.4cm]{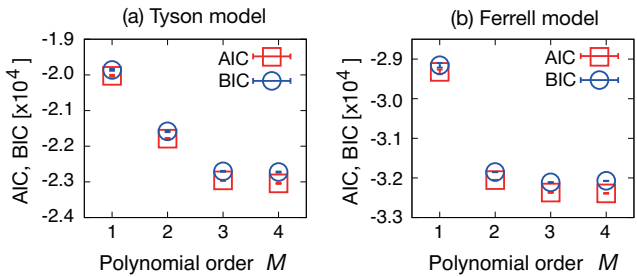}}
  \caption{
    The AIC and BIC are plotted against different polynomial orders $M=1,2,3,4$
    for the (a) Tyson and (b) Ferrell models.
    For each evaluation, 20 different initial parameters are sampled to avoid the local minima.
    After the learning, the average and standard deviation are calculated from 100 evaluation trials based
    on the particle filter.
  }
  \label{fig:ic}
\end{figure}

The simplest polynomial form required for reproducing the observed dynamics
is determined by starting with linear equations composed of active Cdc2 and Cyclin
(system dimension $D=2$ and polynomial order $M=1$)
and increasing the $D$ and $M$ by one.
It turns out that $D=2$ is sufficient for reproducing a given time-series data set as shown below.
The polynomial order $M$ is determined by minimizing the information criteria
through an optimization of the balance between the goodness of fit and the model complexity
\cite{Akaike1974a, Schwarz1978}.
The Akaike information criterion (AIC) and Bayesian information criterion (BIC)
 are evaluated from the log likelihood, parameter number, and data size
 for each model (Fig.\ref{fig:ic}).
Both the AIC and BIC show a decrease from $M=1$ to $3$, but an increase or insignificant decrease at $M=4$.
Therefore, we analyze models with $D = 2$ and $M = 3$
 (see the section 3 in the Supplemental Material for the learned parameter values and detailed settings in the learning algorithm).

\begin{figure}[tbp]
  \centering
  \subfigure[]{\includegraphics[width=8.6cm]{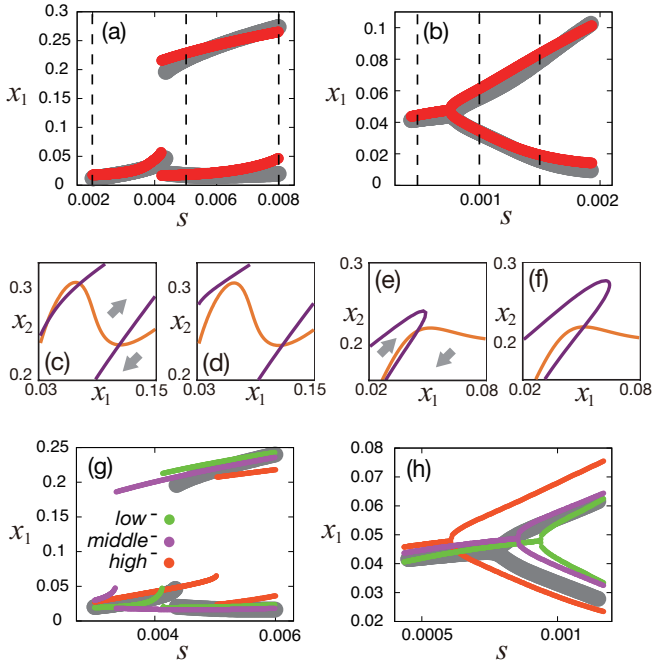}}
  \caption{
    Bifurcation diagrams of the (a) Tyson and (b) Ferrell models.
    The active Cdc2 concentration $x_1$ for the learned systems is plotted against the Cyclin synthesis rate $s$ (red),
    and corresponding concentrations of the data generators are also shown for comparison (gray).
    The broken lines indicate points at which the data are given.
    For the Tyson model, there is another attractor with a tiny basin that is ignored.
    (c-f) The nullclines of the learned systems around the bifurcation points are shown.
    Purple and orange lines represent nullclines of $x_1$ (active Cdc2) and $x_2$ (Cyclin), respectively,
    and the gray arrows indicate the flow direction.
    (c,d) The learned system from the Tyson model exhibits SNIC and
    (e,f) that from the Ferrell model exhibits a supercritical Hopf bifurcation.
    The values of the bifurcation parameter are
    (c) $s = 0.0038$, (d) $0.0044$, (e) $0.0005$, and (f) $0.0012$, respectively.
    (g,h) Bifurcation diagrams using the data at two of the three bifurcation parameter points.
    The learning that lacks data at the lowest, intermediate, and highest bifurcation parameter values
    are denoted by as $low^-$, $middle^-$ and $high^-$, respectively.
  }
  \label{fig:bif}
\end{figure}

To check whether the learning procedure can capture the bifurcation of the original
 data generator system, we compare the bifurcation diagrams of the learned systems
 with those of the data generators.
Figures \ref{fig:bif}(a) and (b) show bifurcation diagrams against Cyclin synthesis rate $s$ (red lines)
 for the learned systems in the Tyson and Ferrell models, respectively.
The bifurcation diagrams for the corresponding noiseless data generators are shown
 by the gray lines.
Although the data for the learning are given only at three bifurcation parameter points
 (indicated by the broken lines), the learned systems have quantitatively
 similar diagrams to those of the corresponding data generators.
The sudden appearance of a limit cycle with finite amplitude is reproduced for the Tyson model,
while the gradual increase in amplitude from the bifurcation point is reproduced for
the Ferrell model.
These features are characteristics of the SNIC and supercritical Hopf bifurcation.
Nullclines of the learned systems in the vicinity of the bifurcation points are shown
in Figs. \ref{fig:bif}(c) and (d) for the Tyson model and in Figs. \ref{fig:bif}(e) and (f)
for the Ferrell model.
The results confirm the onset of SNIC and supercritical Hopf bifurcation, respectively.
Thus, each learned system inherits the bifurcation type of the original model
 through the learning procedure in spite of noisy and partial observations.

When the learning is conducted by using the data on two of the three bifurcation
 parameter points, the learned systems still exhibit the correct bifurcation types,
 although the points of oscillation onset and amplitudes are biased (Figs. \ref{fig:bif}(g) and (h)).
Note that identification of bifurcation is possible even by using the data only on
 one side of a bifurcation point (as indicated by the green lines).
These results indicate the interesting possibility that
 the learning procedure can predict the type of bifurcation that will occur from the
 data before the bifurcation point only.

\begin{figure}[tbp]
  \centering
    \subfigure[]{\includegraphics[width=8cm]{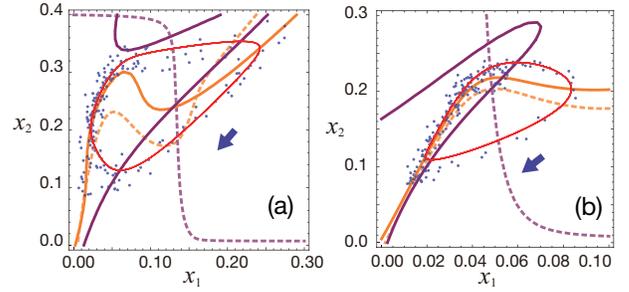}}
  \caption{
    Comparison of the learned systems and reduced models for the (a) Tyson and (b) Ferrell models.
    The purple and orange lines represent the nullclines of $x_1$ (active Cdc2) and $x_2$ (Cyclin), respectively,
    for the learned systems (solid lines) and reduced models (broken lines).
    A noisy time series from the data generators (blue dots) and the orbits of the learned models (red lines)
    are also shown.
    The blue arrows indicate the flow direction.
    The values of the bifurcation parameter are (a) $s = 0.006$, (b) $0.0015$.
  }
  \label{fig:compare}
\end{figure}

We also show here how the high-dimensional phase space structures of the original data generators
 are mapped onto the lower-dimensional surfaces in the learned systems.
Reduced two-variable models are derived by adiabatic elimination
 following a similar procedure by Novak and Tyson \cite{Novak1993a}
(see the section 4 in the Supplemental Material for the detailed procedure and reduced model equations).
Like the learned systems, the reduced models are composed of active Cdc2 and total Cyclin.
Figure \ref{fig:compare} shows the nullclines of the learned systems (the solid orange and purple lines)
 and the reduced models (the broken lines).
In both the Tyson and Ferrell models, the learned system and reduced model nullclines for active Cdc2 have a similar $N$-shaped form (orange lines),
indicating the existence of positive feedback in the molecular circuits.
In contrast, those for the total Cyclin disagree quite significantly.
To check the consistency of the nullclines and dynamics, Fig. \ref{fig:compare} also shows a noisy
 time series from the data generators (blue points) and the orbit of the learned system (red lines).
The nullclines of the learned systems are consistent with the dynamics in the data
 but the reduced models are not.
This failure arises because
 the dynamics of a component mediating the inhibition from active Cdc2 to Cyclin is not fast
enough to allow the adiabatic approximation.
Higher-order contribution beyond the adiabatic elimination performed here should be
 included, which requires complicated technical work.
Nevertheless, the learning process automatically reproduces the appropriate
 low-dimensional dynamics and estimates the bifurcation structures without
 knowledge of the detailed high-dimensional model systems.

Gathering biological data is complicated by intrinsic and observation noises,
 partial observation, and a small number of possible experimental conditions.
We have outlined here a machine-learning procedure
 based on likelihood maximization that makes use of all the information in the time-series data,
 including that in the noise.
By using synthetic data that share the difficulties found in actual biological data,
we demonstrated that the procedure could derive low-dimensional model
 equations that reproduced the obtained time-series data and
captured the bifurcation types of the original systems.
These results support the conjecture that the learning procedure will be able to
construct reliable low-dimensional models for real time-series data of active Cdc2 and Cyclin
 levels in future studies.
Being able to identify the model systems and bifurcation types will provide a useful method
for elucidating both the molecular interactions in the circuit and the biological functions of the dynamics.
Further, since the dynamics and bifurcation are found widely among various biological processes,
 the method is expected to be applicable to various cell system with cell-imaging data.

We note that the proposed procedure can be interpreted as a reduction method from
 high- to low-dimensional systems like adiabatic approximation.
In particular, in the vicinity of the bifurcation points, the systems are usually
 reduced to normal forms represented by low-dimensional differential equations
 with low-order polynomial forms \cite{Guckenheimer1983}.
 However, unlike analytical reduction methods that
 require the original high-dimensional equations, the present learning procedure
 uses only the time-series data.
This is especially advantageous for studying cell dynamics that involve
 complex molecular interactions.
On the other hand, since the learning method has a less theoretical basis
 for interpreting the obtained equations, it should be complemented by an analytical procedure.

In essence, the proposed method performs quantitative inference of the phase space
 structures of the dynamical systems.
Therefore, not only the bifurcation structure but also other properties of the dynamical system
 can be analyzed using the same theoretical groundwork developed here.
The detection of phase sensitivity from noisy data
in studies of biological clocks \cite{Winfree2001} would provide an interesting application.
In addition, the method is flexible enough to be combined with other machine-learning techniques;
it was recently shown that compressive sensing exhibits a high performance for learning dynamical systems
\cite{Wang2011a}.
Those possible extensions will further improve our method depending on the situations and experimental setup.
In summary, the proposed method will be an efficient way to capture the essential features of
the cellular dynamics by mediating dynamical system modeling with experimental observations.

\begin{acknowledgments}
  We would like to thank K. Kamino, N. Saito, and S. Sawai for illuminating comments and stimulating discussions.
  This work was supported by the Grant-in-Aid MEXT/JSPS (No. 24115503).
\end{acknowledgments}
\bibliographystyle{apsrev}

\end{document}


\begin{center}
\textbf{\Large{
Supplemental Material: Identifying dynamical systems with bifurcations from noisy partial observation}}

~\\

Yohei Kondo$^*$, Kunihiko Kaneko, Shuji Ishihara

Graduate School of Arts and Sciences, University of Tokyo, 3-8-1 Komaba, Meguro-ku, Tokyo 153-8902, Japan
\end{center}
{\small * kondo@complex.c.u-tokyo.ac.jp}

\section{ Learning algorithm}
\subsection{State space model}
We first introduce the state space model composed of the state
equations and observation equations describing the system dynamics and
observation process, respectively.  Let us consider $D$-dimensional
stochastic differential equations that describe a system, and $d$
components in the system that are observed simultaneously.  In the
model, the state variable $x_i (i=1,\dots,D)$ evolves under the
function $f_i(\{x_j\},s)$, where $s$ represents an input to the
system, and the observation value $y_i$ is obtained through the
function $g_i(\{x_j\})$.

By discretizing the dynamics in time with the Euler-Maruyama scheme
\cite{Kloeden}, we can write the space state model as
\begin{eqnarray}
  x_i^{t+1} &=& x_i^t + \Delta t f_i(\{x_j^t\},s) + \sigma_i \xi_i^t\sqrt{\Delta t},\\
  y_i^r &=& g_i(\{x_j^r\}) + \eta_i \phi_i^r,
\end{eqnarray}
where $t (\in T)$ and $r (\in R)$ are time points, $\Delta t$ is an
integration time, and $\sigma_i$ and $\eta_i$ are the noise
intensities in the dynamics and observation, respectively.  Both
$\xi_i^t$ and $\phi_i^r$ are sampled from a standard normal
distribution.  In general, the set of observed time points $R$ is a
subset of the entire time point set $T$ (i.e., $R \subseteq T$) for
the numerical integration.  We assume that the function $f_i$ can be
expressed by the summation of linearly independent functions $(f_i^n
(n=1,\dots,N_i))$ as
\begin{eqnarray}
  f_i(\{x_j^t\},s) = \sum_{n=1}^{N_i} k_i^n f_i^n(\{x_j^t\},s).
\end{eqnarray}
Here, $\{k_i^n\}$ are the coefficients to be estimated.

Let us consider that $A$ time-series data sets are given.  The
learning procedure estimates the parameters $\{k_i^n\}$,
$\{\sigma_i\}$, $\{\eta_i\}$, and all the true states $\{x_i^t\}$ for
each time-series set.  In our method, the initial condition for the
$i$th component in the $a$th time-series set is assumed to obey a
Gaussian distribution parameterized by the mean $\mu_{i,a}$ and the
variance $V_{i,a}$.  Distributions at other points are automatically
estimated by the particle filter algorithm explained below.  Then, the
parameters to be estimated are $\theta =
(\{k_i^n\},\{\sigma_i\},\{\eta_i\},\{\mu_{i,a}\},\{V_{i,a}\})$.

\subsection{SAEM algorithm}
Our aim is to find model parameters $\theta$ by maximizing log
likelihood function
\begin{eqnarray}
  \log L(\theta) = \log p(Y|\theta) = \log \int_{X}p(Y|X,\theta)p(X|\theta)\label{suppeq:logl}.
\end{eqnarray}
Here, $Y$ $( = \{Y_a\}, a=1, \dots, A)$ denotes the data sets of the
$A$ time series, and $X$ $( = \{X_a\}, a=1, \dots, A)$ denotes the
entire time series of estimated states.  We employ an EM algorithm
that maximizes $\log P(X,Y|\theta)$ (the complete-data log likelihood
function), which is equivalent to maximizing the likelihood in
eq. (\ref{suppeq:logl})\cite{Dempster1977}.  By iterating two steps known as the E and M
steps, the states $X$ and parameters $\theta$ are estimated
alternately.  Since our implement of the E step includes the
Monte-Carlo method as described below, stochastic approximation EM
(SAEM) algorithm is adopted \cite{Delyon1999}.  The SAEM procedure is
described as follows.

\vspace{5mm}
\break
\noindent\fbox{SAEM} ~ -------------------------------------------------------------------------------------------------------------
\begin{enumerate}
  \item Initialize the parameter vector $\theta = \theta_0$, and set
    the iteration number $I$ to zero.
  \item (E step) Calculate the posterior distribution of the entire
    time series of state variable $p(X|Y,\theta)$.
  \item (M step) Rename $\theta$ as $\theta_{\rm old}$, and update the
    parameter vector as
    \begin{eqnarray}
      \theta_{\rm new} = \argmax_{\theta} Q_I(\theta)
    \end{eqnarray}
    \begin{eqnarray}
      Q_I = \left\{
      \begin{array}{ll}
        Q(\theta,\theta_{\rm old}) & (I = 0),\\
        (1-\alpha_I)Q_{I-1}(\theta) + \alpha_I Q(\theta,\theta_{\rm old}) & (I > 0),
      \end{array}
      \right.
    \end{eqnarray}
    where
    \begin{eqnarray}
      Q(\theta,\theta_{\rm old}) = \langle\log p(X,Y|\theta)\rangle_{p(X|Y,\theta_{\rm old})},
    \end{eqnarray}
    and $\{\alpha_I\}$ is a non-increasing sequence of positive values converging to zero.
  \item Increment $I$ by one, and iterate steps 2 and 3 until the estimation of the parameter vector converges.
\end{enumerate}
~~~------------------------------------------------------------------------------------------------------------------------

\vspace{5mm}
The details of the E and M steps are described in the following
sections.

\vspace{5mm}
\subsection{E step}
Since different time-series data are independent stochastic variables, we can write
\begin{eqnarray}
  \log p(X|Y,\theta) = \sum_a \log p(X_a|Y_a,\theta).
\end{eqnarray}
Then, each $\log p(X_a|Y_a,\theta)$ is evaluated by using a particle
filter algorithm that approximates the non-Gaussian distribution of
the state $x_i^t$ as a collection of many particles, each of which
represents a sample from the distribution
\cite{Gordon1993,Kitagawa1996}.  Specifically, the algorithm required
here is called a particle smoother.  For the $a$th time series, let
$x_{i,a}^{t,p}$ denote the $p$th particle for representing
$x_{i}^{t}$, and let $y_{i,a}^t$ denote an observed value at time $t$.
The procedure of the particle smoother is described as follows.

\vspace{5mm}
\noindent\fbox{Particle smoother} ~ ----------------------------------------------------------------------------------------\\
For $a = 1,\dots, A$
\begin{enumerate}
  \item For $i = 1,\dots, D$ and $p = 1,\dots, P$,
    set the initial states as $x_{i,a}^{0,p} \sim N(\mu_{i,a},V_{i,a})$,
    and normalize the weights as $w_{a}^{p} = 1/P$
  \item At each $t = 0,1,\dots$
    \begin{itemize}
      \item For $i = 1,\dots, D$ and $p = 1,\dots, P$, calculate $x_{i,a}^{t+1,p}$ from $x_{i,a}^{t,p}$
        by using the state equations.
      \item If $t = r \in R$, update the weights of the particles as
        \begin{eqnarray}
          w_{a}^{p} &=& \frac{w_{a,{\rm old}}^{p}l_{a}^{r,p}}{\sum_p
            w_{a, {\rm old}}^{p}l_{a}^{r,p}},
        \end{eqnarray}
        where
        \begin{eqnarray}
          l_{a}^{r,p} = \prod_{i}^{d}p(y_{i,a}^{r}|\{x_{j,a}^{r,p}\}).
        \end{eqnarray}
      \item If $P_{eff} = 1/\sum_p(w_a^p)^2 < P_{thres}$ (i.e., if effective number of the particles fall below
        a threshold value), resample the
        particles according to the new weights.  Note that the history
        of particles $(x_{i,a}^{0,p}, x_{i,a}^{1,p}, \dots,
        x_{i,a}^{r-1,p})$ is resampled in parallel.
    \end{itemize}
  \item Finish when all data points have passed ($t = \max (R)$), and
    estimate the log likelihood as
    \begin{eqnarray}
      \log L_a(\theta) = \sum_{r\in R} \log (\frac{1}{P}\sum_p^P l_{a}^{r,p}).
    \end{eqnarray}
\end{enumerate}
~~~------------------------------------------------------------------------------------------------------------------------

\vspace{5mm}
Then, the posterior distribution of the time series of the state is
approximated as
\begin{eqnarray}
  p(X_a|Y_a,\theta_{\rm old}) = \sum_{p=1}^{P} w_a^{p}\delta(X_a - X_a^p),
\end{eqnarray}
where $X_a^p$ indicates a sample path ($\{x_{i,a}^{t,p}\}, i =
1,\dots,D, t \in T$).  On the basis of this approximation, we
calculate the average of the complete-data log likelihood as
\begin{eqnarray}
 Q(\theta,\theta_{\rm old})
  &=& \langle\log p(\{X_a\},\{Y_a\}|\theta)\rangle_{p(\{X_a\}|\{Y_a\},\theta_{\rm old})}\\
  &=& \sum_{a=1}^A \sum_{p=1}^P w_a^p \log p(X_a^p,Y_a|\theta)\\
  &=&
  \sum_{a=1}^A \sum_{p=1}^P \sum_{i=1}^{D} w_a^p\left( -\frac{1}{2}\log 2\pi V_{i,a}
  - \frac{(x_{i,a}^{0,p} - \mu_{i,a})^2}{2V_{i,a}} \right)\\ \nonumber
  &+& \sum_{a=1}^A \sum_{p=1}^{P} \sum_{t\in T} \sum_{i=1}^D
  w_a^p \left( -\frac{1}{2}\log 2\pi (\sigma_i)^2
  - \frac{\left( (x_{i,a}^{t+1,p} - x_{i,a}^{t,p})
    - \Delta t \sum_{n=1}^{N_i} k_i^n f_i^n(\{x_{j,a}^{t,p}\}) \right)^2}{2(\sigma_i)^2\Delta t}\right)\\ 
  &+& \sum_{a=1}^{A} \sum_{p=1}^{P} \sum_{r \in R} \sum_{i=1}^d
  w_a^p \left( -\frac{1}{2}\log 2\pi (\eta_i)^2 - \frac{(y_{i,a}^{r,p} - g_j(\{x_{j,a}^{r,p}\}))^2}{2(\eta_j)^2} \right). \nonumber
\end{eqnarray}

\vspace{5mm}
\subsection{M step}
At the $I$th iteration, the parameter-value update is performed by
finding the $\theta$ for which $\frac{d}{d\theta}Q_I(\theta) = 0$.  We
describe the case of $\frac{d}{d\theta}Q(\theta,\theta_{\rm old}) = 0$
for simplicity, although the optimization problem can be solved
generally by the same method.  The following example demonstrates the
determination of parameters of the system dynamics ($k_i^n$) and the
strength of the system noise ($\sigma_i$) in detail.

First, by differentiating the complete-data log likelihood with
respect to $k_l^m$, we obtain
\begin{eqnarray}
  0 &=& \pdif{}{k_l^m}Q(\theta,\theta_{\rm old})\\
  &=& \sum_a^A \sum_p^P \sum_{t\in T} w_a^p \left(
  \frac{\left((x_{l,a}^{t+1,p} - x_{l,a}^{t,p}) - \Delta t \sum_n^{N_l} k_l^n f_l^n(\{x_{j,a}^{t,p}\})\right)\Delta t f_l^m(\{x_{j,a}^{t,p}\})}
       {2(\sigma_l)^2\Delta t}
       \right) \nonumber\\
\nonumber   &=& \sum_{a,p,t} w_a^p (\Delta x_{l,a}^{t,p}f_l^m(\{x_{j,a}^{t,p}\}))
       - (\sum_n^{N_l} k_l^n \sum_{a,p,t}w_a^p f_l^n(\{x_{j,a}^{t,p}\})f_l^m(\{x_{j,a}^{t,p}\})),
\end{eqnarray}
where $\Delta x_{l,a}^{t,p} = (x_{l,a}^{t+1,p} - x_{l,a}^{t,p})/\Delta
t$.  By defining the vectors $b_l$ and $k_l$ and a matrix $A_l$ as
\begin{eqnarray}
  (b_l)_m &=& \sum_{a,p,t}w_a^p (\Delta x_{l,a}^{t,p}f_l^m(\{x_{j,a}^{t,p}\}))\\
  (k_l)_m &=& k_l^m\\
  (A_l)_{nm} &=& \sum_{a,p,t}w_a^p f_l^n(\{x_{j,a}^{t,p}\})f_l^m(\{x_{j,a}^{t,p}\}),
\end{eqnarray}
the system dynamics parameters are determined as follows.
\begin{eqnarray}
  k_l = (A_l)^{-1}b_l.
\end{eqnarray}

Next, using the new $k_i^n$ calculated above, we obtain
\begin{eqnarray}
  \pdif{}{\sigma_l}Q(\theta,\theta_{\rm old}) 
  = \sum_{a,p,t}w_a^p\left(-\frac{1}{\sigma_l} +
  \frac{\Delta t^2\left( \Delta x_{l,a}^{t,p} - \sum_n^{N_l} k_l^n f_l^n(\{x_{j,a}^{t,p}\})\right)}{(\sigma_l)^3\Delta t}\right) = 0,
\end{eqnarray}
and thus,
\begin{eqnarray}
  (\sigma_l)^2 = \Delta t \sum_{a,p,t} w_a^p\left( \Delta x_{l,a}^{t,p} - \sum_n^{N_l} k_l^nf_l^n(\{x_{j,a}^{t,p}\})\right)^2.
\end{eqnarray}

The other parameters are estimated in the same manner.  Only for the
variance of the initial condition $V_{i,a}$, we define the minimum
value $V_{min}$ to avoid an unnaturally small value resulting from a
problem called sample impoverishment \cite{Arulampalam2002}.

\vspace{15mm}
\section{Data preparation}
The model equations and parameter values used in the present study as
the data generators are described here.  Each data generators, the
Tyson and Ferrell models, is numerically integrated with white
Gaussian noise by using a stochastic Runge-Kutta II (SRKII) algorithm
\cite{Honeycutt1992}.  To prohibit negative values of the chemical
concentrations as a result of noise, each variable is reset to a small
positive value $\epsilon$ (= 0.001) when the value is less than
$\epsilon$.  We confirmed that the results of the present study are
stable so long as $\epsilon$ is sufficiently small.

\vspace{5mm}

The Tyson model described in Novak \& Tyson \cite{Novak1993}.
{\small
\begin{eqnarray}
  \frac{d{\rm [cyclin]}}{dt} &=& k_{1}[{\rm AA}] - k_{2}[{\rm cyclin}] - k_3[{\rm Cdc2}][{\rm cyclin}] + \xi_1 (t)\\
  \frac{d\ccn}{dt} &=& k_{PP} \cc{tp} - (k_{wee} + k_{cak} + k_2)\ccn\\ \nonumber
  &+& k_{cdc25}\cc{yp} + k_3[{\rm Cdc2}][{\rm cyclin}] + \xi_2 (t)\\
  \frac{d \cc{yp}}{dt} &=& k_{wee}\ccn - (k_{cdc25} + k_{cak} + k_2)\cc{yp}\\ \nonumber
  &+& k_{PP}\cc{yptp} + \xi_3 (t)\\
  \frac{d \cc{yptp}}{dt} &=& k_{wee}\cc{tp} - (k_{PP} + k_{cdc25} + k_2)\cc{yptp}\\ \nonumber
  &+& k_{cak}\cc{yp} + \xi_4 (t)
\end{eqnarray}
}
{\small
\begin{eqnarray}
  \frac{d \cc{tp}}{dt} &=& k_{cak}\ccn - (k_{PP} + k_{wee} + k_2)\cc{tp}\\ \nonumber
  &+& k_{cdc25}\cc{yptp} + \xi_5 (t)\\
  \frac{d \cdcnigo}{dt} &=& k_{a}\cc{tp}\frac{(\cdcnigotot - \cdcnigo)}{K_a + (\cdcnigotot - \cdcnigo)}\label{suppeq:tyson_cdc25}\\ \nonumber
  &-& k_{b}[{\rm PPase}]\frac{\cdcnigo}{K_b + \cdcnigo} + \xi_6 (t)\\
  \frac{d \apcs}{dt} &=& k_{c}\iep \frac{(\apctot - \apcs)}{K_c + (\apctot - \apcs)}
  - k_{d}[{\rm anti\mathchar`-IE}]\frac{\apcs}{K_d + \apcs} + \xi_7 (t) \label{suppeq:tyson_apc}\\
  \frac{d \wees}{dt} &=& k_{e}\cc{tp}\frac{(\weetot - \wees)}{K_e + (\weetot - \wees)} \label{suppeq:tyson_wee}\\
  &-& k_{f}[{\rm PPase}]\frac{\wees}{K_f + \wees} + \xi_8 (t) \nonumber\\
  \frac{d \iep}{dt} &=& k_{g}\cc{tp}\frac{(\ietot - \iep)}{K_g + (\ietot - \iep)}
  \label{suppeq:tyson_iep}\\
  &-& k_{h}[{\rm PPase}]\frac{\iep}{K_h + \iep} + \xi_9 (t)\nonumber\\
  k_{cdc25} &=& V_{cdc25'}(\cdcnigotot - \cdcnigo) + V_{cdc25''}\cdcnigo \label{suppeq:tyson_kcdc25}\\
  k_{wee} &=& V_{wee'}\wees + V_{wee''}(\weetot - \wees) \label{suppeq:tyson_kwee}\\
  k_2 &=& V_{2'}(\apctot  - \apcs) + V_{2''}\apcs \label{suppeq:tyson_k2}\\
  \rm{[Cdc2]}&=&
\rm{[Cdc2}_{tot}] - \ccn - \cc{yp} \\
  &-& \cc{yptp} - \cc{tp} \nonumber 
\end{eqnarray}
}
Noise terms are represented by $\xi_i(t)$ with white
Gaussian statistics $ \langle\xi_i (t)\rangle = 0$ and $\langle\xi_i
(t)\xi_j (\tau)\rangle = 2 \sigma_i^2 \delta_{i,j}\delta(t - \tau)$.

\vspace{5mm}
\subsection{Ferrell model}
The Ferrell model is described in Tsai et al. \cite{Tsai2008}.
{\small
\begin{eqnarray}
  \frac{d{\rm [cyclin]}}{dt} &=& k_{synth} - k_{dest}[{\rm APC}^*][{\rm cyclin}] - k_a[{\rm Cdc2}][{\rm cyclin}]
    + k_d\ccn + \xi_1 (t)\\
  \frac{d\ccn}{dt} &=& k_a[{\rm Cdc2}][{\rm cyclin}] - k_d\ccn - k_{dest}\apcs \ccn \\ \nonumber
  &-& k_{wee1}\wees \ccn
  - k_{wee1basal}(\weetot - \wees)\ccn\\ \nonumber
  &+& k_{Cdc25}\cdcnigo \cc{yp}
  + k_{Cdc25basal}(\cdcnigotot - \cdcnigo)\cc{y}\\ \nonumber
  &+& \xi_2(t)\\
  \frac{d \cc{yp}}{dt} &=& k_{wee1}\wees \ccn + k_{wee1basal}(\weetot - \wees)\ccn\\ \nonumber
  &-& k_{Cdc25}\cdcnigo \cc{yp}
  - k_{Cdc25basal}(\cdcnigotot - \cdcnigo)\cc{yp}\\ \nonumber
  &-& k_{cak}\cc{yp} + k_{pp2c}\cc{yptp} - k_{dest}\apcs \cc{yp}\\ \nonumber
  &+& \xi_3(t)\\
  \frac{d \cc{yptp}}{dt} &=& k_{cak}\cc{yp} - k_{pp2c} \cc{yptp}\\ \nonumber
  &-& k_{Cdc25}\cdcnigo \cc{yptp}
  - k_{Cdc25basal}(\cdcnigotot - \cdcnigo) \cc{yptp}\\ \nonumber
  &+& k_{wee1}\wees \cc{tp}
  + k_{wee1basal}(\weetot - \wees) \cc{tp}\\ \nonumber
  &-& k_{dest}\apcs \cc{yptp} + \xi_4(t)
\end{eqnarray}
\begin{eqnarray}
  \frac{d \cc{tp}}{dt} &=& k_{Cdc25}\cdcnigo \cc{yptp} \\ \nonumber
  &+& k_{Cdc25basal}(\cdcnigotot - \cdcnigo)\cc{yptp}\\ \nonumber
  &-& k_{wee1}\wees \cc{tp} - k_{wee1basal}(\weetot - \wees) \cc{tp}\\ \nonumber
  &-& k_{dest}\apcs \cc{tp} + \xi_5(t) \nonumber\\
  \frac{d \cdcnigo}{dt} &=& k_{Cdc25on}\frac{\cc{tp}^{n_{cdc25}}}{{\rm EC50}_{Cdc25}^{n_{cdc25}} + \cc{tp}^{n_{cdc25}}}
  (\cdcnigotot - \cdcnigo) \\\nonumber
  &-& k_{Cdc25off}\cdcnigo + \xi_6(t)\\
  \frac{d \wees}{dt} &=& - k_{Wee1off}\frac{\cc{tp}^{n_{wee1}}}{{\rm EC50}_{wee1}^{n_{wee1}} + \cc{tp}^{n_{wee1}}}\wees \\ \nonumber
   &+& k_{Wee1on}(\weetot - \wees) + \xi_7(t)\\
  \frac{d \plxs}{dt} &=& k_{plxon}\frac{\cc{tp}^{n_{plx}}}{{\rm EC50}_{plx}^{n_{plx}} + \cc{tp}^{n_{plx}}}
  (\plxtot - \plxs) \\
&  - &k_{plxoff}\plxs + \xi_8(t)\nonumber\\
  \frac{d \apcs}{dt} &=& k_{apcon}\frac{\plxs^{n_{apc}}}{{\rm EC50}_{apc}^{n_{apc}} + \plxs^{n_{apc}}}
  (\apctot - \apcs) - k_{apcoff}\apcs + \xi_9(t)
\end{eqnarray}
\begin{eqnarray}
  {\rm [Cdc2]} = {\rm [Cdc2}_{tot}] - \ccn - \cc{yp} - \cc{yptp} 
   - \cc{tp}
\end{eqnarray}
}
Noise terms are represented by $\xi_i(t)$ with white
Gaussian statistics $ \langle\xi_i (t)\rangle = 0$ and $\langle\xi_i
(t)\xi_j (\tau)\rangle = 2 \sigma_i^2 \delta_{i,j}\delta(t - \tau)$.

\vspace{5mm}
\subsection{Artificial measurement process}
For both the Tyson and Ferrell models, the artificial measurement
process is implemented as follows.
\begin{eqnarray}
  y_1 &=& \mbox{active Cdc2} = \frac{\cc{tp} + \eta_1\phi_1}{\rm [Cdc2_{tot}]}\\
  y_2 &=& \mbox{Total Cyclin}\\ \nonumber
  &=& \big(\rm{[cyclin}] + \ccn + \cc{yp} + \cc{yptp}
 \\ &&   + ~\cc{tp} +\eta_2\phi_2\big)/\rm{[Cdc2_{tot}]}. \nonumber
\end{eqnarray}
Here, $\eta_{1,2}$ is the observation noise intensity, and
$\phi_{1,2}$ is sampled from a standard normal distribution.

\vspace{5mm}
\subsection{Parameter values}
The parameters used in the Tyson and Ferrell models are listed in
Tables S\ref{table:tyson} and S\ref{table:ferrell}, respectively.  In
the Tyson model, all the parameter values are the same as those in the
original literature \cite{Novak1993}.  In the Ferrell model, all the
parameter values are the same as those in Pomerening et
al. \cite{Pomerening2005} except for ``$factor$," which is set as in
\cite{Tsai2008}.

\subsection{Orbits of the data generators}
\begin{figure}[H]
  \centering
  \subfigure[]{\includegraphics[width=6cm]{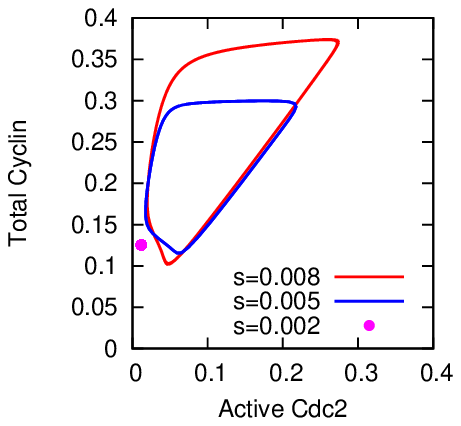}}
  \subfigure[]{\includegraphics[width=6cm]{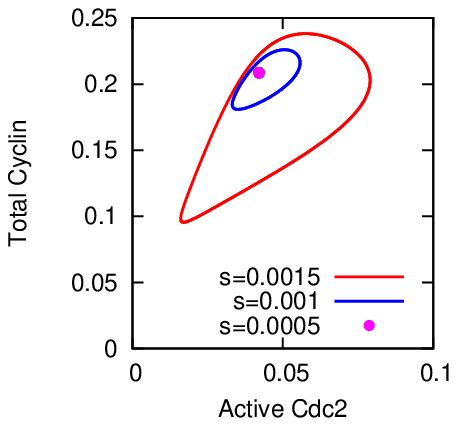}}
  \caption{
    Time series generated from the (a) Tyson and (b) Ferrell models
    through the artificial measurement process in the case that
    both system and observation noises are zero.}
  \label{suppfig:orbit}
\end{figure}

Figure S\ref{suppfig:orbit} show noiseless orbits of the data generators.
We note that each orbit exhibits no intersection, indicating that
the two observable variables are sufficient to abstract the original high-dimensional dynamics.

\subsection{Data set}
The data used for the learning are shown in supplemental Fig. S\ref{suppfig:data}.

\vspace{15mm}
\section{Settings and results of learning}
Parameters in the learning algorithm is listed in Table
S\ref{table:learning_parameter}. The learned parameters of the
third-order polynomial model used in the main text is shown in Table
S\ref{table:model_parameter}.

\vspace{15mm}
\section{Model reduction}
We reduce the Tyson and Ferrell models to two-dimensional systems by
the same procedure as described in \cite{Novak1993a}.  By denoting the
non-dimensionalized active Cdc2 and total Cyclin levels as $u$ and
$v$, respectively, the reduced models are written as follows.

The reduced Tyson model takes the form as
\begin{eqnarray}
  \dot{u} &=& \frac{s}{1 + k_{PP}/k_{cak}} - (f_{\rm APC}(u)
  + f_{\rm Wee}(u)) u + f_{\rm Cdc25}(u) (\frac{v}{1 + k_{PP}/k_{cak}} - u),\\
  \dot{v} &=& s - f_{\rm APC}(u) v,
\end{eqnarray}
where $f_{\rm Cdc25}, f_{\rm Wee},$ and $f_{\rm APC}$ are the
functions corresponding to the adiabatic solutions of
eqs. (\ref{suppeq:tyson_kcdc25}), (\ref{suppeq:tyson_kwee}), and
(\ref{suppeq:tyson_k2}), respectively.

This reduction procedure includes the determination of the level of
Cdc2-Cyclin-tp (i.e., the value of $u$) from the sum $[{\rm
    Cdc2\mathchar`-Cyclin}] + [{\rm
    Cdc2\mathchar`-Cyclin\mathchar`-tp}]$ (see Appendix A in
\cite{Novak1993a}).  This is based on a detailed balance assumption of
the phosphorylation reaction between the two molecular species.
However, in the Ferrell model, the absence of the reaction makes the
original reduction procedure inapplicable.  Then, we simply assume $
[{\rm Cdc2\mathchar`-Cyclin\mathchar`-tp}] \sim [{\rm
    Cdc2\mathchar`-Cyclin}] + [{\rm
    Cdc2\mathchar`-Cyclin\mathchar`-tp}]$, because it is observed that
the ratio of $[{\rm Cdc2\mathchar`-Cyclin}]$ to the summation is small
throughout the dynamics within the parameter region we consider.
Consequently, the reduced Ferrell model is written as
\begin{eqnarray}
  \dot{u} &=& s - (f_{\rm APC}(u)) + f_{\rm Wee}(u)) u + f_{\rm Cdc25}(v - u),\\
  \dot{v} &=& s - f_{\rm APC}(u) v,
\end{eqnarray}
where $f_{\rm Cdc25}, f_{\rm Wee}$, and $f_{\rm APC}$ are derived from
the adiabatic approximation in the same manner as the Tyson model.

\newpage

\vspace{3mm}
\begin{table}[H]
\begin{minipage}[t]{.5\textwidth}
  \centering
  \small
  \begin{tabular}{p{.59\textwidth}p{.15\textwidth}}
    \hline
    \textit{Parameter}&\\ \hline\hline
   ~~$k_1$[AA]/[Cdc2$_{tot}$]&$s$\\
   ~~$k_3$[Cdc2$_{tot}$]&1.0\\
   ~~$k_{CAK}$&0.25\\
   ~~$k_{PP}$&0.025\\
   ~~$V_{2'}\apctot$&0.015\\
   ~~$V_{2''}\apctot$&1.0\\
   ~~$V_{cdc25'}\cdcnigotot$&0.1\\
   ~~$V_{cdc25''}\cdcnigotot$&2.0\\
   ~~$V_{wee'}\weetot$&0.1\\
   ~~$V_{wee''}\weetot$&1.0\\ \hline
   ~~$k_a$[Cdc2$_{tot}$]/$\cdcnigotot$&1.0\\
   ~~$k_b$[PPase]/$\cdcnigotot$&0.125\\
   ~~$k_c$$\ietot/\apctot$&0.1\\
   ~~$k_d$[anti IE]/$\apctot$&0.095\\
   ~~$k_e$[Cdc2$_{tot}$]/$\weetot$&1.33\\
   ~~$k_f$[PPase]/$\weetot$&0.1\\
   ~~$k_g$[Cdc2$_{tot}$]/$\ietot$&0.65\\
   ~~$k_h$[PPase]/$\ietot$&0.087\\ \hline
   ~~$K_a$/$\cdcnigotot$&0.1\\
   ~~$K_b$/$\cdcnigotot$&0.1\\
   ~~$K_c$/$\apctot$&0.01\\
   ~~$K_d$/$\apctot$&0.01\\
   ~~$K_e$/$\weetot$&0.3\\
   ~~$K_f$/$\weetot$&0.3\\
   ~~$K_g$/$\ietot$&0.01\\
   ~~$K_h$/$\ietot$&0.01\\ \hline
   ~~$\sigma_{1,\dots,5}$/[Cdc2$_{tot}$]&0.0024\\
   ~~$\sigma_6$/[Cdc25$_{tot}$]&0.0024\\
   ~~$\sigma_7$/[APC$_{tot}$]&0.0024\\
   ~~$\sigma_8$/[Wee$_{tot}$]&0.0024\\
   ~~$\sigma_9$/[IEP$_{tot}$]&0.0024\\
   ~~$\eta_1$/[Cdc2$_{tot}$]&0.005\\
   ~~$\eta_2$/[Cdc2$_{tot}$]&0.005\\ \hline
\\
\\
\end{tabular}
\caption{Parameters in the Tyson model}
\label{table:tyson}
\end{minipage}
\hfill
\begin{minipage}[t]{.5\textwidth}
  \centering
  \small
  \begin{tabular}{p{.45\textwidth}p{.28\textwidth}}
    \hline
    \textit{Parameter}&\\ \hline\hline
   ~~$k_{syn}$&$s$\\
   ~~$k_{dest}$&0.006\\
   ~~$k_a$&0.1\\
   ~~$k_d$&0.001\\
   ~~$factor$&10\\
   ~~$k_{wee1}$&0.05\\
   ~~$k_{wee1basal}$&$k_{wee1}/factor$\\
   ~~$k_{Cdc25}$&0.1\\
   ~~$k_{Cdc25basal}$&$k_{cdc25}/factor$\\
   ~~$k_{cak}$&0.8\\
   ~~$k_{pp2c}$&0.008\\
   ~~$k_{Cdc25on}$&1.75\\
   ~~$k_{Cdc25off}$&0.2\\
   ~~$k_{wee1on}$&0.2\\
   ~~$k_{wee1off}$&1.75\\
   ~~$k_{plxon}$&1.0\\
   ~~$k_{plxoff}$&0.15\\
   ~~$k_{apcon}$&1.0\\
   ~~$k_{apcoff}$&0.15\\ \hline
   ~~EC50$_{Cdc25}$&40\\
   ~~EC50$_{wee1}$&40\\
   ~~EC50$_{plx}$&40\\
   ~~EC50$_{apc}$&40\\ \hline
   ~~$n_{cdc25}$&4\\
   ~~$n_{wee1}$&4\\
   ~~$n_{plx}$&3\\
   ~~$n_{apc}$&3\\ \hline
   ~~Cdc25$_{tot}$&15\\
   ~~Wee1$_{tot}$&15\\
   ~~Plx$_{tot}$&50\\
   ~~APC$_{tot}$&50\\
   ~~Cdc2$_{tot}$&230\\
    \hline
   ~~$\sigma_{1,\dots,9}$&0.06\\
   ~~$\eta_1$&0.4\\
   ~~$\eta_2$&0.4\\ \hline
  \end{tabular}
  \caption{Parameters in the Ferrell model}
  \label{table:ferrell}
\end{minipage}
\end{table}

\vspace{6mm}
\begin{table}[H]
  \small
  \centering
  \begin{tabular}{ll}
    \hline
    \textit{Parameters in the learning algorithm} &\\
    \hline\hline
    ~~Number of time series $A$ &~~ 9\\
    ~~Integration time $\Delta t$ &~~ 1.0\\
    ~~Entire-time points $T$  &~~$\{0,1,2,\dots, 400\}$\\
    ~~Observed-time points $R$ &~~ $\{0,2,4,\dots, 400\}$\\
    ~~Decreasing sequence in SAEM $\alpha_I$&~~ 1 $(I \leq 30)$,
$1/\sqrt{(I-30)}$ (otherwise)\\
    ~~Particle number $P$ &~~ 1000\\
    ~~Particle number threshold for resampling $P_{thres}$ &~~ 500\\
    ~~Initial system dynamics parameters $\{k_i^n\}$ &~~ uniform
distribution [-0.001,0.001]\\
    ~~Initial system noise strength $\sigma_i$ &~~ 0.1\\
    ~~Initial observation noise strength $\eta_i$&~~ 0.2\\
    ~~Minimum value of initial condition variance $V_{min}$&~~ 0.001\\
    \hline
  \end{tabular}
  \caption{Pre-determined parameters used in the learning algorithm.}
  \label{table:learning_parameter}
\end{table}

\newpage
\vspace*{10mm}

\begin{table}[H]
  \centering
  \small
  \begin{tabular}{ccc}
    \hline
    ~\textit{Parameter}~~~ & \textit{Learned (Tyson model)} &
~~~\textit{Learned (Ferrell model)}~~\\ \hline\hline
   $k_1^1$ & 0.00064 & -0.00032\\
   $k_1^2$ & -0.334 & -0.343\\
   $k_1^3$ & 0.0655 & 0.0657\\
   $k_1^4$ & -0.8 & 1.05\\
   $k_1^5$ & 2.39 & -1.25\\
   $k_1^6$ & -0.639 & 0.107\\
   $k_1^7$ & -25.3 & -47.3\\
   $k_1^8$ & 31.1 & 47.8\\
   $k_1^9$ & -13.1 & -3.97\\
   $k_1^{10}$ & 2.22 & -0.157\\ \hline
   $k_2^2$ & -0.4 & -0.57\\
   $k_2^3$ & 0.0612 & 0.0637\\
   $k_2^4$ & -4.62 & -0.9\\
   $k_2^5$ & 5.47 & 2.41\\
   $k_2^6$ & -0.805 & 0.000818\\
   $k_2^7$ & -9 & 61.4\\
   $k_2^8$ & 22.7 & -61.3\\
   $k_2^9$ & -15.3 & 18.3\\
   $k_2^9$ & 1.92 & -2.73\\ \hline
   $\sigma_1$ & 0.00646 & 0.00199\\
   $\sigma_2$ & 0.00703 & 0.00205\\ \hline
   $\eta_1$ & 0.00334 & 0.00101\\
   $\eta_2$ & 0.00621 & 0.00133\\
    \hline
  \end{tabular}
  \caption{Learned parameters of the third-order polynomial systems
for the Tyson and Ferrell models.}
  \label{table:model_parameter}
\end{table}

\vspace{20mm}

\newpage
\vspace*{20mm}

\begin{figure}[H]
  \centering
  \subfigure[]{\includegraphics[width=5cm]{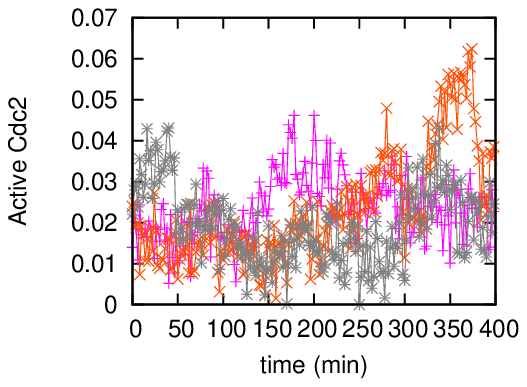}}
  \subfigure[]{\includegraphics[width=5cm]{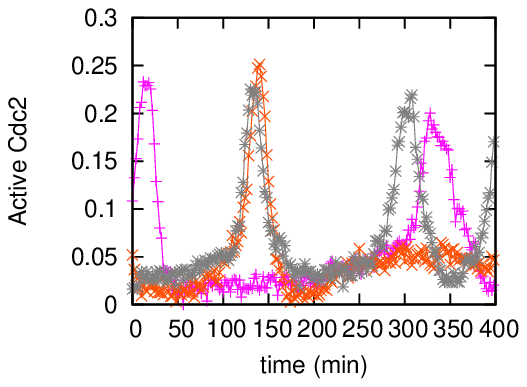}}
  \subfigure[]{\includegraphics[width=5cm]{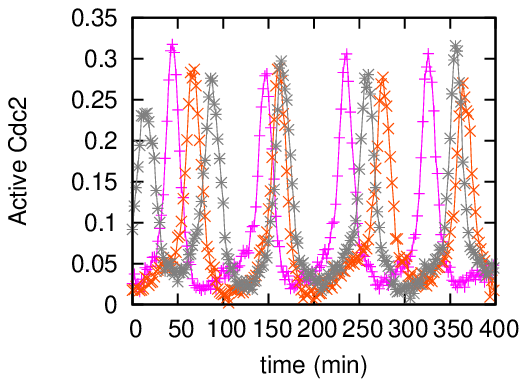}}
  \subfigure[]{\includegraphics[width=5cm]{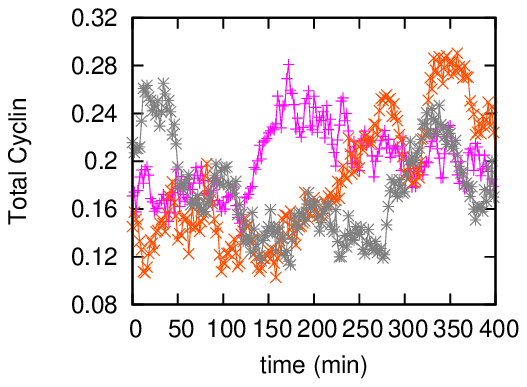}}
  \subfigure[]{\includegraphics[width=5cm]{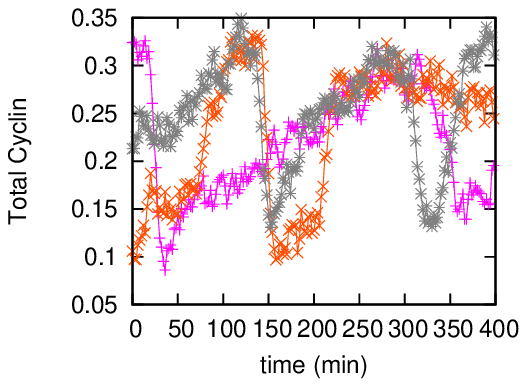}}
  \subfigure[]{\includegraphics[width=5cm]{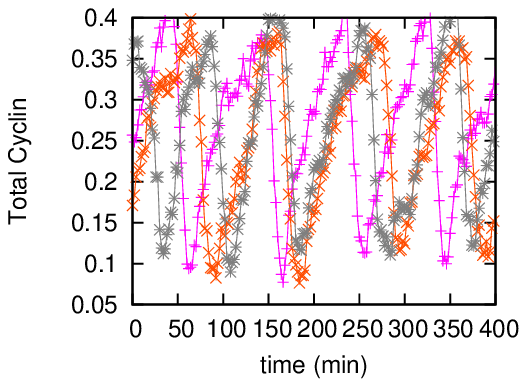}}
  \subfigure[]{\includegraphics[width=5cm]{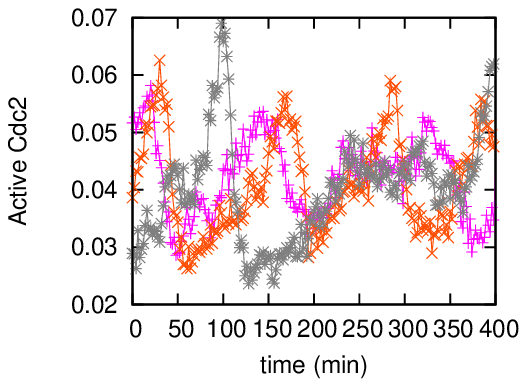}}
  \subfigure[]{\includegraphics[width=5cm]{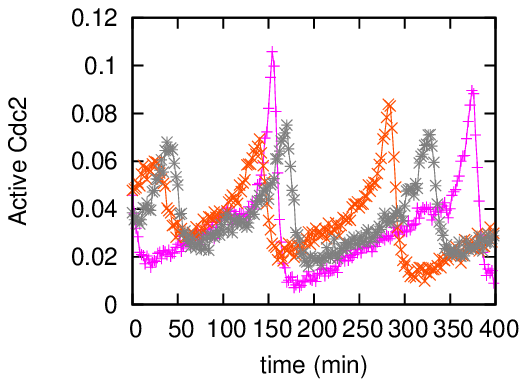}}
  \subfigure[]{\includegraphics[width=5cm]{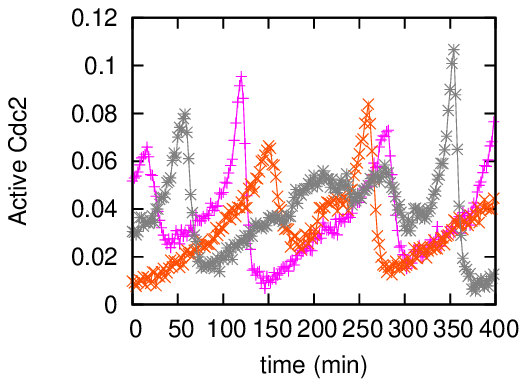}}
  \subfigure[]{\includegraphics[width=5cm]{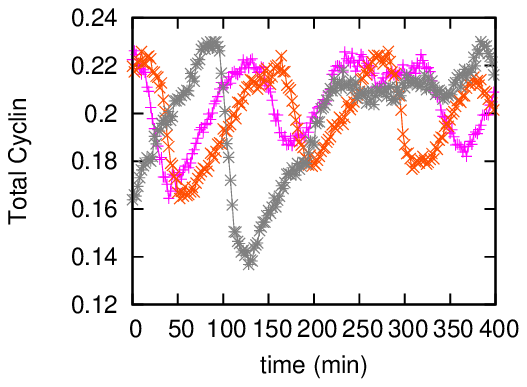}}
  \subfigure[]{\includegraphics[width=5cm]{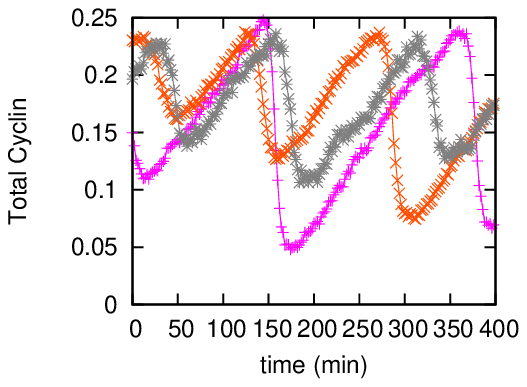}}
  \subfigure[]{\includegraphics[width=5cm]{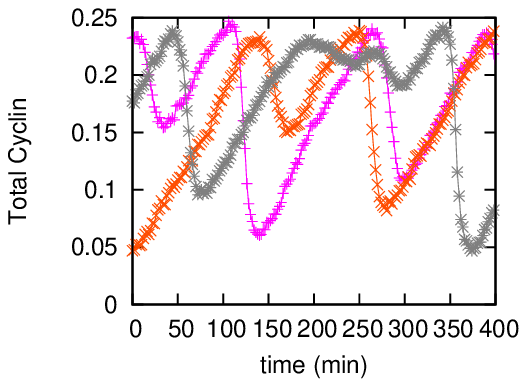}}
  \caption{
    Noisy time series generated from the (a-f) Tyson and (g-l) Ferrell models
    through the artificial measurement process.
    The colors indicate results from independent trials.
    The values of the bifurcation parameter are (a,d) $s = 0.002$, (b,e) $0.005$, (c,f) $0.008$,
    (g,j) $0.005$, (h,k) $0.001$, and (i,l) $0.0015$, respectively.
  }
  \label{suppfig:data}
\end{figure}

\newpage

\bibliographystyle{unsrt}